\documentclass[preprint,12pt,3p]{elsarticle}
\usepackage{multirow}
\usepackage{epsfig} 
\usepackage{graphicx}
\usepackage{soul}
\usepackage{color}
\usepackage{booktabs}
\usepackage{amsmath}
\usepackage{algorithm}
\usepackage{algorithmic}
\usepackage{array}
\usepackage{subfigure}
\usepackage{framed} 
\usepackage{multicol} 
\usepackage{ulem} 
\usepackage[title]{appendix}
\usepackage{lscape}
\usepackage{hyperref}
\usepackage[dvipsnames]{xcolor}
\usepackage{ifthen}
			
\usepackage{nomencl} 
\makenomenclature
		
\usepackage{enumitem}	
\setlist{nolistsep} 

\usepackage{amssymb}
\usepackage{lineno}

\newcommand{\cyr}[1]{{\color{black} #1}}       

\journal{Renewable Energy}

\begin{document}

\begin{frontmatter}

\title{ Benchmarks for  \cyr{Solar Radiation} Time Series \cyr{Forecasting}}

\author[label1]{Cyril Voyant}\corref{cor1} 
\author[label1]{Gilles Notton}\corref{cor1}
\author[label1]{Jean-Laurent Duchaud}\corref{cor1}
\author[label1]{Luis Antonio García Gutiérrez}\corref{cor1} 
\author[label2]{Jamie M. Bright}
\author[label3]{Dazhi Yang}\corref{cor1} 
\address[label1]{University of Corsica, SPE Laboratory, France}
\address[label2]{UK Power Networks, London, UK}
\address[label3]{Harbin Institute of Technology, School of Electrical Engineering and Automation, Harbin, Heilongjiang, China}

\begin{abstract}
With an ever-increasing share of intermittent renewable energy in the world's energy mix, \cyr{there is an increasing need for advanced solar power forecasting models to optimize the operation and control of solar power plants}. In order to justify the need for more elaborate forecast modeling, one must compare the performance of advanced models with na\"ive reference methods. On this point, \cyr{a rigorous formalism using statistical tools, variational calculation and quantification of noise in the measurement is studied and} five na\"ive reference forecasting methods are considered, among which there is a newly proposed approach called ARTU (a particular autoregressive model of order two). These methods do not require any training phase nor demand any (or almost no) historical data. Additionally, motivated by the well-known benefits of ensemble forecasting, a combination of these models is considered, and then validated using data from multiple sites with diverse climatological characteristics, based on various error metrics, among which some are rarely used in the field of solar energy. The most appropriate benchmarking method depends on the salient features of the variable being forecast (e.g., seasonality, cyclicity, or conditional heteoroscedasity) as well as the forecast horizon. Hence, to ensure a fair benchmarking, forecasters should endeavor to discover the most appropriate na\"ive reference method for their setup by testing all available options. \cyr{Among the methods proposed in this paper, the combination and ARTU statistically offer the best results for the proposed study conditions.}


\end{abstract}


\begin{highlights}
\item Benchmark of six \cyr{Statistical Reference Methods (SRM)} 
\item \cyr{Direct multi-step forecast strategy without training phase} 
\item Validation of results using data from multiple climates 
\item \cyr{Theory mixing statistical tools, variational calculation and measurement error }
\item \cyr{Combination of models and ARTU are the best performing models}
\end{highlights}

\begin{keyword}
 Irradiation\sep  Filtering \sep Exponential Smoothing \sep Combination \sep Benchmark \sep  Forecast


\end{keyword}

\end{frontmatter}


\section{Introduction}
\label{sec:1}

Leveraging physics-based and statistical methods to explain and forecast time series has been playing an increasingly important role in energy meteorology \citep{article3,NOTTON201896}. It is logical to expect a rapid development of forecasting methods, in terms of both number and intricacy \citep{makridakis2021future}. In order to justify the choice of one method over another, forecast comparison is necessary. In that, superiority claims ought to be regarded as an essential component in all forecasting works. Under the ``publish or perish'' regime, overarching statements such as ``the proposed method outperforms the existing ones'' are ubiquitous and rapidly updated, which tends to result in a literature that is difficult to follow, to interpret, and to condense. Precautions must be taken. 

According to Wolpert's ``no free lunch'' theorem \citep{inbook1}, each time a predictive method is deemed superior in some aspects, there ought to be some other aspects in which the method does not perform as well comparing to its alternatives. In this regard, one may consider the classic example given by \citet{Gneiting2011}, in which two sets of forecasts optimized under different objectives (least squares and least absolute deviations) outperform each other under different evaluation metrics (root mean square error and mean absolute error); this is formally known as \textit{consistency in forecast verification} \citep{YANG202020}. In a more general sense, even when a simple method is compared to a sophisticated method, the former method has the advantage of being easy to implement, although its accuracy may not be as attractive as the latter. On this point, the problem of choosing a forecasting method, like many other problems of a similar nature, is one of balance. 

Another important aspect of forecast comparison is the situation under which the verification is conducted. In the field of renewable energy, in particular, intermittent solar energy, the optimality of a method depends on the geographical location, climate and weather regime \citep{Bright2019markov}, time-step, and forecast horizon; it does not make much sense to compare directly two sets of forecasts issued under distinct forecasting situations. Suppose some forecasts have been reported to be optimal in one forecasting situation, it may not compare favorably against its peers in another situation \citep{Yang2021predictabiliy}. Stated differently, knowing the optimal choice of method to be used in a particular situation is not straightforward, and one cannot simply take as true the conclusions made elsewhere. This is a limitation as to the general validity of inductive reasoning. In that, the only logical way to confirm that a method is indeed optimal is to test all existing methods for that given forecast situation.

This task is a necessarily difficult one. It is for that reason that, until now, there is no consensus on what constitutes a perfect model \citep{elke,FLIESS2018519,SOUBDHAN2016246}. Since enumerating and testing all existing methods is never quite possible, it is customary to leverage a na\"ive reference method during forecasting, such that forecasts made at different locations and over different time periods can be compared, though with appropriate caveats. Depending on the amount of improvement acquired, there would be different levels of enthusiasm for the methods of interest. 
In order for this approach to take effect, most generally, several prerequisites must be evaluated carefully. Firstly, reference methods should be applicable in various independent and operational modes \citep{DAVID2021672,en13143565,doi:10.1063/1.5114985,YANG2019410}. That is to say that the reference method should be ``universal'' in a sense that it does not depend on the type of available data. Secondly, the reference method must be sufficiently na\"ive, in that, it does not require dedicated knowledge to implement. For this reason, simple or very simple reference methods are preferred over elaborate reference methods that only provide slightly better results. Thirdly, when multiple na\"ive reference methods are present, the one with the highest accuracy should be used \citep{murphy_climatology_1992}. 

There are many reference methods in weather forecasting \citep{HONG2014357,9218967}. The most widely used ones are climatology and persistence \citep{YANG2019981} though the accuracy of these two references are generally low. Other time series reference methods are occasionally used, such as the simple exponential smoothing \citep{hyndman_forecasting_2018}. If we were to expand the list of reference methods in that direction, the choice becomes more flexible. In this paper, we propose to explore further options in order to effectively judge the quality of any new or existing forecasting methods \cyr{using time series foramlism} in the solar radiation field. A new baseline approach is introduced to have a global indicator of the performance of all the methods. \cyr{Among these methods, there are regressive methods, stochastic learning methods, deep learning methods, genetic algorithm, data-driven methods, local-sensing methods, hybrid forecasting methods, and application orientated methods.}

The following pages are organized as follows. In Section \ref{sec:2}, various reference forecasting methods are introduced, which include four well-known methods, a newly proposed one, and a linear combination of the previous five. In Section \ref{sec:experiment}, the specifics related to the forecasting of climate and weather time series are underlined. In Section \ref{sec:result}, an application of these methods is firstly demonstrated on global horizontal irradiance (GHI) data measured in Ajaccio, France. Then, we check if the conclusions obtained for this site can be generalized for GHI series from four other sites with different climatological characteristics. The forecasting methods are applied to tilted global irradiance (TGI), which is more useful for solar photovoltaic applications, in view to verify the relevance of the methods. The final results refer to the temperature and wind time series in order to judge the portability of the methods we want to qualify as reference. Finally, Section \ref{sec:conclusion} is dedicated to conclusions. 

\cyr{As is discussed further in the next section, it is crucial to develop the notion of statistical reference method (SRM) for solar radiation forecasting. It is in this context, that we revisit classical methods that are either widely or not widely used resulting in two new proposed methods based on a novel formalism.}

\section{\cyr{Statistical Reference Methods (SRM)}}
\label{sec:2}

The existing methodologies that are most used in the context of solar resources forecasting, to facilitate integration and management of  \cyr{photovoltaic (PV)} plants, smart grid control, or energy trading on the spot market, are based on the time series formalism and numerical weather prediction (NWP). Comparing these two classes of methods, the former is based on extrapolation of data, whereas the latter solves the governing partial differential equations which describes the state of the atmosphere. Since the time series approaches are far less computational demanding than NWP, they are more attractive in terms of forecast benchmarking. In this paper, all benchmarking methods considered are in accordance with time series approaches; they are abbreviated as statistical reference methods (SRMs).

Considering a signal $x$ with samples $x_t$ being regularly-spaced values in time, and transforming it such that $x^c_t\equiv x_t-\mathbb{E}[x]$, which implies the transformed variable has expected value: $\mathbb{E}[x^c]=0$. One fundamental way to describe such a signal is to assume the forecast is a function of the most recent observation(s), or an autoregressive (AR) model. For a forecast at horizon $h$ and a direct multi-step forecast strategy\footnote{to contrast with recursive multi-step forecast \citep{Bontempi2013} where the prediction for the prior time step is used as an input for making a prediction on the following time step}, it is expressed:

\begin{equation}
\label{eq:eq1}
{x}_{t+h}^c=\alpha x^c_t+\omega_{t+h},
\end{equation}
where $x^c_t$ is a centered time-dependant signal, $\alpha$ ($\vert \alpha \vert<1$ for process stability) is a gain term and $\omega$ is additive noise. The direct strategy could be applied because in solar forecasting context this is often the best method although there is no real consensus on this subject. As mentioned by \citet {Taieb2012RecursiveAD}, ``Choosing between these different strategies involves a trade-off between bias and estimation variance". Under the state space framework in the discrete domain where Kalman filter is often used, $x^c_t$ could be defined as the state vector of the process at time $t$, and $\alpha$ the state transition parameter of the process from the state at $t$ to the state at $t+h$, assuming stationarity over time. $\omega$ is the associated zero-mean white noise process, i.e.\, $\omega\sim \mathcal{N}(0, \sigma^2_{\omega})$ with $\sigma^2_{\omega}=\mathbb{E}[\omega_t^2]$. 

With this in mind, observations on this variable are contained in the series $y_t$. The centered form, denoted $y^c_t$ can be modeled by the observation equation:

\begin{equation}
\label{eq:eq2}
y^c_{t}\equiv y_t-\mathbb{E}(y)=x^c_t+v_{t},
\end{equation}
where $y_t$ is the actual measurement of $x$ at time $t$ and $v_t$ is the associated measurement error with $v\sim\mathcal{N}(0,\sigma^2_{v})$. Note that the hypothesis $\mathbb{E}(x)=\mathbb{E}(y)\equiv \bar{y}$ is implied in Eq.(\ref{eq:eq2}) and therefore, a bias in the measurement would be detrimental: a quality control of the sensors is a prohibitive prerequisite. 

\subsection{Usual Benchmark Methods for Meteorological Time Series}
There are many reference methods in meteorology and mainly in solar radiation. In this section and in \ref{appendix}, we will list the most commonly used ones.  

\subsubsection{Persistence \textnormal{(PER)}}
\label{sec:subsec1}
Persistence is the simplest case where $\alpha=1$ in Eq.(\ref{eq:eq1}), the prediction of $x^c_{t+h}$ denoted $\widehat{x}^c_{t+h}$ becomes:

\begin{equation}
\label{eq:eq3}
\widehat{x}^c_{t+h}=x^c_t
\end{equation}
which implies that $\widehat{x}_{t+h}=y_t$. This model (see \ref{algo:Pers} for details) is very reliable when the  meteorological data series has a very low variability but it quickly becomes ineffective with very noisy, periodic or trending signals. However, in  operational cases (forecasting for energy or power management), when the user needs a forecast and when there is no measurement history, it is often the only way to proceed in the nowcasting case (from 1h to 6h). For benchmarking at larger horizons, a solution (and undoubtedly the best alternative) would be the use of climatology. 

\subsubsection{Climatology \textnormal{(CLIM)}}
\label{sec:subsec2}
If in the previous case the method used was only valid for strongly persistent phenomenona (low variability), the model presented here corresponds to the case where there is no statistical dependence between the different measurements (i.e.\ white noise). In this case, we should choose $\alpha= 0$ where it immediately becomes $\widehat{x}^c_{t+h}=0$ and therefore $\widehat{x}_{t+h}^{\,}=\mathbb{E}(y)$ (historical mean). This model (see \ref{algo:Clim} for details), though simple, should not be overlooked, because as soon as the autocorrelation coefficient $\rho$ comes close to 0, it is often the best way to attain a minimum forecast error. This model becomes interesting when predicting meteorological time series, and when deep horizons close to the predictability limit of chaotic phenomena are studied. 

\subsubsection{ \cyr{Autoregressive Model of Order One \textnormal{AR(1)}} or Climatology Persistence \textnormal{(CLIPER)}}
\label{sec:AR}
In this section the link between the AR(1) estimate and the well-named ``climatology--persistence combination,'' or CLIPER for short, is shown. This kind of model can be considered as a reference because it is easy to implement and does not require any learning phase. Hence, it is easy to use and robust in case of data are suddenly missing due to a detector failure if the forecasts are used automatically in operational mode. Now, consider the Yule--Walker equations \citep{175742} or equivalently multiply Eq.(\ref{eq:eq1}) by $x^c_t$ and take the expectation value. $x^c_t$ is considered as a stationary process in the weak sense in the rest of the paper (also denoted wide-sense stationarity, or covariance stationarity). It follows that:

\begin{equation}
\label{eq:eq4bis}
\mathbb{E}[x^c_{t+h}x^c_{t}]=\mathbb{E}[\alpha x^c_tx^c_t+\omega_{t+h}x^c_t].
\end{equation}

After some mathematical considerations detailed in \ref{appendix:cliper} and using the correlation factor $\rho(h)$, it comes that the prediction can be expressed by:

\begin{equation}
\label{eq:CLIPER}
\widehat{x}_{t+h}=\rho(h)y_t+\left[1-\rho(h)\right]\mathbb{E}(y).
\end{equation}

An alternative, and shorter, way to arrive at Eq.~(\ref{eq:CLIPER}) can be found in \citep{doi:10.1063/1.5114985}, in which the same results can be obtained by minimizing the mean square error (MSE) of CLIPER forecasts. The pseudo-code used to predict with CLIPER is given in \ref{algo:ClimPers}.

\subsubsection{Simple Exponential Smoothing \textnormal{(ES)}}
\label{sec:ETS}
Forecasts are calculated using weights that decrease exponentially as data become older (where $\vert \alpha \vert <1$ is the smoothing parameter) as described in Eq.(\ref{eq:ES}) \citep{hyndman_forecasting_2018}.

\begin{equation}
\label{eq:ES}
    \widehat{x}^c_{t+h}=\alpha x^c_t+\alpha(1-\alpha)x^c_{t-1}+\alpha(1-\alpha)^2x^c_{t-2}+...
\end{equation}
That is, all forecasts take the same value ($h$-step-ahead forecast is constant $\forall h$), equal to the last level component. Remember that these forecasts are only  suitable if the time series has no trend or seasonal component. Applying the same tools exposed in the beginning of Section \ref{sec:2}, Eq.(\ref{eq:ES}) can be replaced with:

\begin{equation}
\label{eq:eq7bis}
    \widehat{x}_{t+h}-\bar{y}=\alpha( y_t-\bar{y})+\alpha(1-\alpha)(y_{t-1}-\bar{y})+\alpha(1-\alpha)^2(y_{t-2}-\bar{y})+...
\end{equation}

The separation of the $\bar{y}$ and $y_{t-i}$ terms leads to:

\begin{equation}
\label{eq:eq7ter}
    \widehat{x}_{t+h}=\bar{y}[1-\alpha-\alpha(1-\alpha)-\alpha(1-\alpha)^2-...] + \alpha y_t+\alpha(1-\alpha)y_{t-1}+\alpha(1-\alpha)^2y_{t-2}+...
\end{equation}

If the second part of this equation (Eq.(\ref{eq:eq7ter})) can be reduced according to the sum $\sum_{i=0}^{n} \alpha(1-\alpha)^i y_{t-i}$, the first term is related to the sum of the first $n+1$ terms of a geometric series (common ratio = $1-\alpha$). In the end, the ES model is given by: 

\begin{equation}
\label{eq:eq7}
    \widehat{x}_{t+h}=\alpha \sum_{i=0}^{n} (1-\alpha)^i y_{t-i} +\bar{y}(1-\alpha)^{n+1},\alpha \neq 0.
\end{equation}

In the exponential smoothing method (see \ref{algo:ES} for details), the predictions lie between the two extremes, the naive persistence $\widehat{x}_{t+h}=y_t$ (when $\alpha=1$) and the simple average $\widehat{x}_{t+h}=\bar{y}$ (when $\alpha=0$), which assumes that all observations are of equal importance, and assigns them equal weights when generating forecasts. In order to be consistent with other methods, we choose to determine $\alpha$ without using an optimization phase. As such, considering that $\alpha=\rho(1)$ we obtain a predictor close to the persistence when $\rho(1)$ is close to 1 and close to the mean when it tends to 0. Note that we could use the relation $ \alpha = \rho (h) $ and this could make sense but we wanted to present the simplest method relating to exponential smoothing.  

\subsection{Proposed Methodologies}
The methods used as reference or so-called naive methods in terms of weather forecasting through the time series formalism are often (and this is the entirely the purpose) very easy to implement. Qualifying the following methods as naive may \cyr{seem  counterintuitive}  given the calculations required to develop it. However, they are only ever performed once, then methods are applicable very simply to whatever the studied time series (solar radiation or not).

\subsubsection{Particular  \cyr{Autoregressive Model of Order Two }  \textnormal{(ARTU)}}
\label{sec:method}
In this part, we propose a methodology that improves the performances of the AR(1) model (or CLIPER) previously discussed in Section (\ref{sec:AR}). We show that, even if this is not part of the initial hypotheses [see Eq.(\ref{eq:minus}-\ref{eq:eq8})], this approach is equivalent to an AR(2), with the difference being that the estimation of the coefficients is only based on the autocorrelations generation---since the method is a particular form of AR(2), it is called ARTU, which is pronounced as ``A-R-two.'' Firstly, it is useful to think of another version of Eq.(\ref{eq:eq1}), in which the state estimate is known to be \textit{not} optimal $({x^{c-}_{t+h}})$.

\begin{equation}
\label{eq:minus}
{x^{c-}_{t+h}}=\alpha x^c_t+\omega_{t+h}.
\end{equation}

We further assume the need for an update ($\widehat{x}^c_{t+h}$) related to a linear combination between $\widehat{x}^{c-}_{t+h}$ ($=\alpha x^c_t$) and the previous innovation (or residual i.e. $y^c_{t}-\widehat{x}^{c-}_{t}$). Shown in Eq.(\ref{eq:eq8}) is the updated form, whereby the factor $K$ can be treated as a gain, and the combination as a filtering process. 

\begin{equation}
\label{eq:eq8}
\widehat{x}^c_{t+h}=\widehat{x}^{c-}_{t+h}+K(y^c_{t}-\widehat{x}^{c-}_{t}).
\end{equation}

Note that this approach, although it is quite close to what is observed with prediction of a classical ARMA(1,1) model \citep{CHATFIELD1988411} or a Kalman filtering \citep{DEGOOIJER2006443}, is different. In the first case, $\omega_{t+h}$  would be the residual of $\widehat{x}^c_{t}$ and not $\widehat{x}^{c-}_{t}$, moreover $K$ would be multiplied by $(y^c_{t}-\widehat{x^c_{t}})$ in Eq.(\ref{eq:eq8}). The second one would require a modification in Eq.(\ref{eq:eq8}), in that, $K$ would be multiplied by $(y^c_{t+h}-\widehat{x}^{c-}_{t+h})$. The assumptions of stationarity formulated in Section \ref{sec:AR} are maintained. From Eq.(\ref{eq:minus}) and Eq.(\ref{eq:eq8}), the forecast can therefore be put in the following form:

\begin{equation}
\label{eq:eq8bis}
\widehat{x}^c_{t+h}=\alpha x^c_t+K(y^c_t-\alpha x^c_{t-h}).
\end{equation}

The transition between states and measurements is:

\begin{equation}
\label{eq:eq8ter}
\widehat{x}_{t+h}-\bar{y}=\alpha(y_t-\bar{y})+K[y_t-\bar{y}-\alpha(y_{t-h}-\bar{y})], 
\end{equation}

\begin{equation}
\label{eq:eq8b}
\widehat{x}_{t+h}=(K+\alpha)y_t-(K\alpha)y_{t-h}+(1+K\alpha-K-\alpha)\bar{y}.
\end{equation}
A more practical parameterization can be proposed by letting $P=K\alpha$ and $S=K+\alpha$. This allows us to present $\widehat{x}_{t+h}$ as a convex combination of $y_t$, $y_{t-h}$ and $\mathbb{E}(y)$ from the sum $S$ and the product $P$:

\begin{equation}
\label{eq:eq8c}
\widehat{x}_{t+h}=Sy_t-Py_{t-h}+(1+P-S)\bar{y}.
\end{equation}

Using what has just been presented above and in \ref{proof}, the optimal values of $\alpha$ and $K$ can be determined by minimizing the MSE of ARTU forecasts. Based on the results therein, the optimal $\alpha$ and $K$ can be written in functions of a triplet [$R$, $\rho(h)$,$\rho(2h)$], where $\rho(h)$ and $\rho(2h)$ are two correlation coefficients that have already defined previously, and $R=\sigma^2_{v}/\sigma_x^2$ is linked to the quality of the measurement. The ARTU prediction is described by the pseudo-code in \ref{algo:METH}. There are, nevertheless, some hypotheses that have been formulated, especially concerning the stationarity of the studied time series. Indeed, wide-sense stationary assumption (mean and variance are time-independent and autocovariance and autocovariance can be expressed as functions of the time-lag) is necessary to obtain a closed-form in \ref{eq:eq14} allowing to simply propose $S$ and $P$ in Eq.(\ref{eq:eq8c}). This is important to keep in mind, because if we deviate too much from these conditions (that can be qualified as ideal), there is a good chance that the ARTU approach (prediction with Eq.(\ref{eq:eq8b}) according the $\alpha$ and $K$ parameters obtained solving Eq.(\ref{eq:eq14})) does not give satisfactory results. To arrive at an approximately stationary time series in respect to solar radiation, it is customary to use a clear-sky model \citep{SUN2021110087,SUN2019550,BRIGHT2020685}, which is often simplified from the Beer—Lambert relationship (See Section \ref{sec:season}). 
We would see that the non-stationary phenomenon appears during the temperature and wind speed studies (in Section \ref{sec:other}) where there is no simple knowledge model, where using a ratio to trend allows us to get closer to the ideal and stationary case. 

\begin{figure}
\includegraphics[scale=1.05]{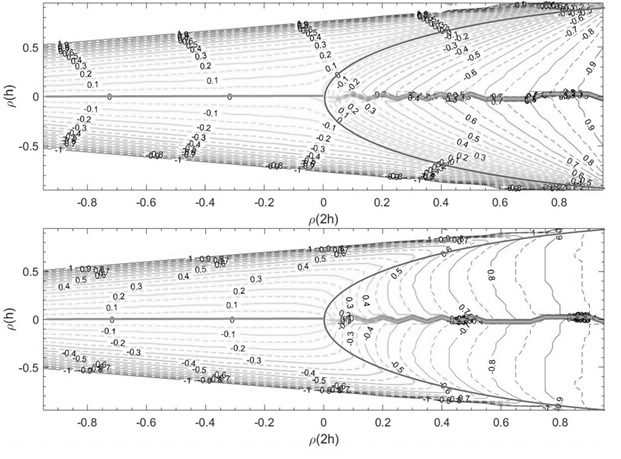}%
\caption{$K$ (top) and $\alpha$ (bottom) values according to the values of $\rho(h)$ and $\rho(2h)$ and for $R=0.05$ (resolution of Eq.(\ref{eq:eq14})).} 
\label{fig:fig1} 
\end{figure}

We computed $\alpha$ and $K$ for values of $\rho(h)$ and $\rho(2h)$ from $-1$ to $+1$ (with an incremental step of 0.01) and for 4 values of $R$ (0, 0.01, 0.05 and 0.1) solving Eq.(\ref{eq:eq14}). An example is given in Fig.\ref{fig:fig1} and other values are available in \url{https://github.com/cyrilvoyant/ARTU.git} (via Matlab\textsuperscript{\tiny\textregistered} codes). What is important to remember is that once these values are known, the forecasts become relatively simple to implement by knowing the autocorrelations of the series studied. The forecast is direct, fast and does not require modeling or learning. The parameters are obtained by data mining approach and the method can then be qualified as na\"ive although a minimum of historical data is necessary. 

A potential limitation on the use of the ARTU approach can be enacted from Eq.(\ref{eq:eq8}). Indeed, by multiplying by $x^c_{t-h}$ and taking the expectation value, it comes Eq.(\ref{eq:eq17}) where we observe a condition between the two correlations of the model.

\begin{equation}
\label{eq:eq17}
\rho(h)^2-\rho(2h)=0.
\end{equation}

This means that predictions can be efficient for autocorrelation factors (ACF) that satisfy this equation but becomes less reliable as soon as we deviate from this case. Note that $\rho(h)=a^{bh}$ is a solution of this equation ($a,b\in\mathbb{R}$ with $ \textnormal{log}(a)b<1$) with the constrain of  $\lvert\rho(h)\rvert<1$. 

As soon as the ACF deviates from the exponential decay, the results are best with the filtering proposed here while when the decay is respected the filtration is not necessary. This induces $K=0$ and $\alpha=\rho(h)$ and the model becomes equivalent to the classical CLIPER presented in Section \ref{sec:AR}. In summary, the best results are observed without filtration and in the presence of the exponential decay of the ACF, but when this one is not observed, the filtration takes all importance and in theory  improves the predictions.

Before going any further, it is thought essential to clarify and discuss the variable $R$. From its construction, we realize that the variance ratio fluctuates between 0 ($ \sigma_x ^ 2  \gg \sigma^2_{v}$) and 1 ($ \sigma_x ^ 2  \sim \sigma^2_{v}$). Usually, the measuring devices used in meteorology are quite efficient, so we can limit the values of $ R $ between $ 0 $ and 10\% of $ \sigma _x ^ 2 $. According to \citet{vuilleumier2017}, one can logically imagine using an $ R $ between 1\% and 5\% . 

\begin{landscape}
\begin{figure}[tb]
\includegraphics[scale=0.80]{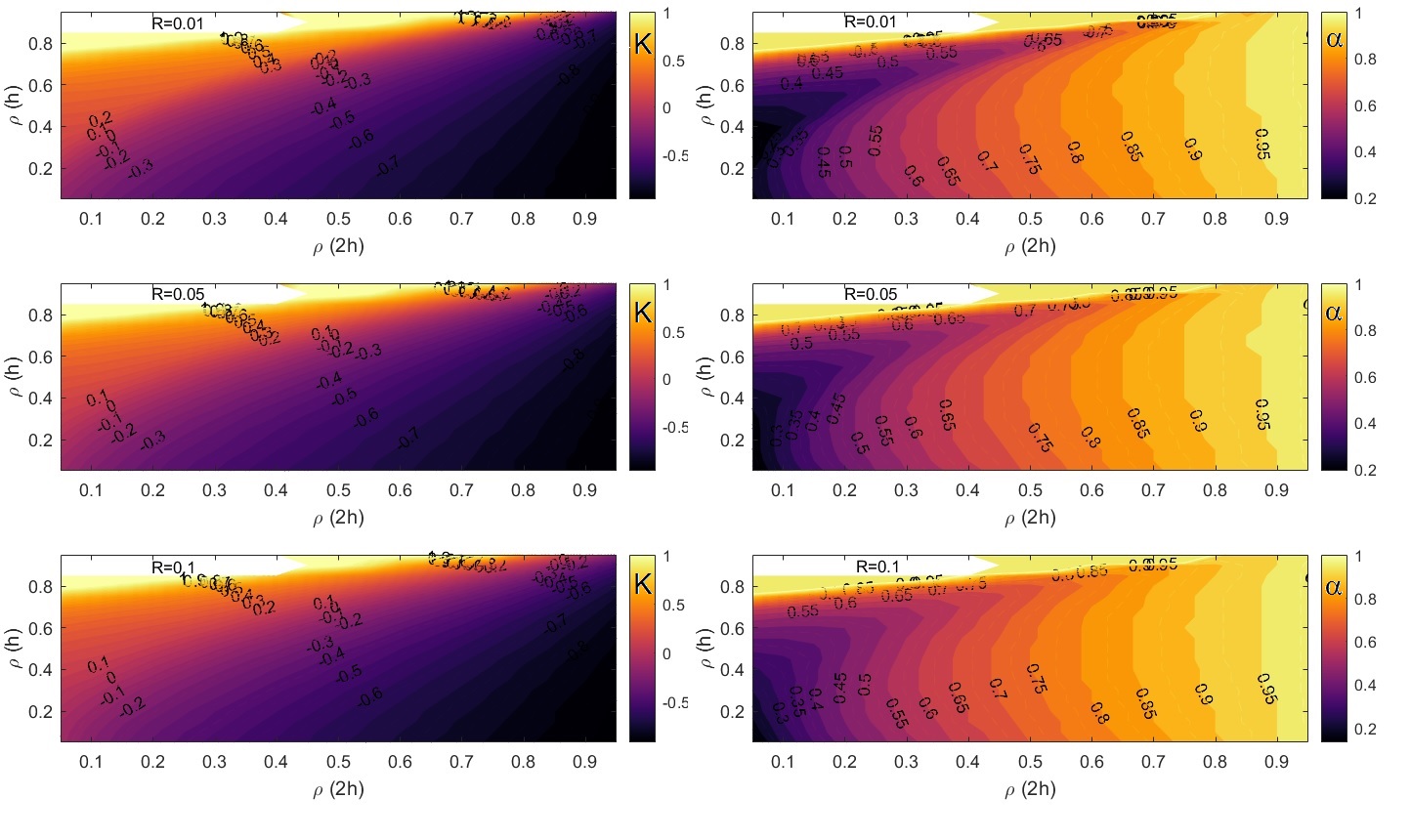}%
\caption{$\alpha$ and $K$ values according to the values of $\rho(h)$, $\rho(2h)$ and $R$} 
\label{fig:fig2}
\end{figure}
\end{landscape}

Fig. \ref{fig:fig1} shows as example (for $R=0.05$) the values of $\alpha$ and $K$ obtained with the methodology exposed previously. Four areas are visible concerning $K$, bounded by the line $\rho(h)=0$ and the parabola defined by the equation $\rho(2h)=\rho(h)^2$. Magnification of the area related to positives value of $\rho(h)$ and $\rho(2h)$ is proposed in Fig. \ref{fig:fig2}. Positive values correspond to what is usually observed in solar radiation. All the files used to generate the $ \alpha $ and $ K $ coefficients are available in \url{https://github.com/cyrilvoyant/ARTU.git} ($\mathcal{M}(R)$ matrices and codes). 
For a forecast at the horizon $h$, the procedure is as follows:
\begin{itemize}[label=$\looparrowright$]  
\item Calculate $ \rho (h) $ and $ \rho (2h) $ using in-sampling data;
\item Obtain the $ \alpha $ and of $ K $ values by consulting on Figures \ref {fig:fig1} and \ref {fig:fig2} or generating them more precisely using the code available in \ref{algo:METH} and the matrices ($\mathcal{M}(R)$); 
\item Derive forecast using Eq.(\ref {eq:eq8c}). 
\end{itemize}
In the end, let us suppose that we seek to use the filtration previously stated, for a signal concerning $h = 1$, if one estimates $ \rho (1) = 0.4 $ and $ \rho (2) = 0.3 $; we therefore have $\widehat{x}_{t+1}=S{y}_t-Py_{t-1}+(1+P-S)\bar{y}$, Cf. Eqs.~(\ref{eq:eq8})--(\ref{eq:eq8c}), and so:

 \begin{itemize}[label=$\looparrowright$]  
\item $\widehat{x}_{t+1}=0.33y_t+0.16y_{t-1}+0.51\bar{y}$ for $R=0.01$, $\alpha=0.60$ and $K=-0.27$ hence $S=0.33$ and $P=-0.16$;
\item $\widehat{x}_{t+1}=0.34y_t+0.15y_{t-1}+0.51\bar{y}$ for $R=0.05$, $\alpha=0.59$ and $K=-0.25$ hence $S=0.34$ and $P=-0.15$;
\item $\widehat{x}_{t+1}=0.35y_t+0.13y_{t-1}+0.52\bar{y}$ for $R=0.10$, $\alpha=0.58$ and $K=-0.23$ hence $S=0.35$ and $P=-0.13$.
\end{itemize}

\subsubsection{ Combination of Methods \textnormal{(COMB)}}
\label{sec:comb}
According to the conclusions related to the M4 forecasting competition \citep{article4}, one of the major findings is related to the combination of methods \citep{SPILIOTIS202159}. In fact, among the most accurate methods, the vast majority (12 of the 17 most accurate) were a combination of statistical approaches. Moreover, this aspect is so important that one innovation was the introduction of a combination reference for benchmarking the accuracy of the methods submitted to the competition. It therefore seems logical to draw inspiration from this remarkable conclusion in order to propose a reference predictive methodology based on the combination of the simple methods presented in this study. Still inspired by \cite{article4} we chose to use the simple arithmetic average of the different outputs of the models although the use of the median may show equally good results in some situations \citep{article5}. \cyr{Note that the term ``combination of methods'' refers to the statistician's universe while machine learning researchers call it ``ensemble learning'' or ``classifier combination''. Whatever the name, this tool is often useful in dealing with potential sources of non-stationary variation present in the data. The Timmermann' paper \citep{TIMMERMANN2006135} allows to understand but especially to validate the use of COMB.}

\section{Experimental setup}
\label{sec:experiment}
To compare the SRMs presented in this paper as objectively as possible, we must establish rules follow best practices in the fields of meteorology, renewable energy, and particularly, solar irradiance forecasting. The seasonality of considered time series and ad-hoc test (based on auto-correlation) are discussed in \ref{sec:season}. In what follows, $I_{GH}$ and $I_{CS}$ are used to denote GHI and its clear-sky expectation, whereas $\widehat{I}_{GH}$ is the GHI forecast. To be more precise, the clear-sky index, which is defined as $\kappa=I_{GH}/I_{CS}$, is used as $x$ as in Eq.~(\ref{eq:eq1}).

\subsection{The pre-treatment}
\label{sec:pre}
As part of this study, several rules and explanations must be given to improve the objectivity of the solar energy forecasting conclusions:
\begin{itemize}[label=$\looparrowright$]
\item Irradiance time series ($I_{GH}$) are measured in several sites around the world with different climatic characteristics. For each of them, the K\"oppen--Geiger classification ($KG$) \citep{ASCENCIOVASQUEZ2019672}, the expected forecastability ($F$) \citep {doi:10.1063/5.0042710} and the geographic coordinates are provided. At least two years of data are available for each site and used during the simulations\cyr{. Usually one year of data is considered sufficient to validate a forecasting method, though two years are preferable;}
\item  The models are evaluated during daytime irradiance values only, filtering the checked data (according to quality control \citep{espinar_quality_2012,articleCQ}) of the solar zenith angle ($I_{GH}$ where zenith angle $\theta_Z>85^\circ$ are excluded);
\item The data was acquired every 15 min or every hour, and the corresponding forecast horizons are between 15 and 150 min, in the first case, and between 1 and 10 h, in the second case. 
\end{itemize}

\subsection{Error Metrics}
In order to compare the accuracy of the forecasting methods, we refer to the official accuracy measures of the M4 competition, i.e., the mean absolute scaled error (MASE \citep{HYNDMAN2006679}) in Eq.(\ref{eq:26})  computed for seasonal time series in retrospective case. MASE admits an average value as denominator, so as long as the filtering of the night hours is operated (see previous subsection), the denominator in Eq.(\ref{eq:26}) will be neither equal nor close to 0. This metric should not be used alone, it is only computed to improve and validate results obtained with the usual methods (Eqs.(\ref{eq:nmae}-\ref{eq:nrmse}) \citep{YANG202020}). Error is calculated on the out-sample data and averaged over all horizons, the main interest of MASE as defined in Eq.(\ref{eq:26}) is tied with that one coefficient for all horizons (retrospective case with multiple step ahead forecasts). 

\begin{equation}
\label{eq:26}
    \textnormal{MASE}\simeq \frac{100}{h}\frac{\sum_{t\in \textnormal{Test}}\sum_{i=1}^{h}\lvert I_{GH}(t+i)-\widehat{I}_{GH}(t+i)\lvert}{\sum_{t\in \textnormal{Test}}\lvert I_{GH}(t)-I_{GH}(t-m)\lvert},
\end{equation}
where $m$ is the period, see Eq.~(\ref{eq:24},and $\textnormal{Test}$ indicates the test sample of size $n$ with $n\gg m$. The MASE method requires a normalization, performed here using the signal period. The filtration presented previously (Section \ref{sec:pre}) has an adverse effect where it reduces the number of exploitable data but also modifies the seasonality (not constant during the year). As a result, it is possible that the denominator of Eq.~(\ref{eq:26}) takes quite high values, which may result in quite low MASE. As all methods are evaluated identically, the interpretation of MASE remains valid, besides a non-periodic version could have been used.

If these metrics are the most commonly used by researchers working on time series forecasting, the analysis of the literature shows that they are only rarely (if ever) used to validate predictions related to global radiation or photo-voltaic power. There are simpler ones that are used in meteorology and more particularly in deterministic solar resource forecasting \citep{YANG202020}, i.e., the normalized mean absolute error (nMAE), see Eq.~(\ref{eq:nmae}) and the normalized root mean square error (nRMSE), see Eq.~(\ref{eq:nrmse}) \citep{metrics}.

\begin{equation}
\label{eq:nmae}
    \textnormal{nMAE}(h)=100\frac{\sum_{t \in \textnormal{Test}}\lvert I_{GH}(t+h)-\widehat{I}_{GH}(t+h)\lvert}{\sum_{t \in \textnormal{Test}} I_{GH}(t)},
\end{equation}

\begin{equation}
\label{eq:nrmse}
    \textnormal{nRMSE}(h)=100\sqrt{n}\frac{\sqrt{\sum_{t \in \textnormal{Test}}[ I_{GH}(t+h)-\widehat{I}_{GH}(t+h)]^2}}{\sum_{t \in \textnormal{Test}} I_{GH}(t)}.
\end{equation}
Keeping in mind that ARTU was built from the minimization of the $L^2$ norm error function (MSE), it would seem logical in the following that the contribution of the filtering is more consistent with nRMSE than with nMAE but it is imperative to quantify what is happening for both.

\section{Results}
\label{sec:result}

All simulations were performed with using Matlab and were run on a standard personal computer (Intel core i7, 16GB RAM). Reports of the execution times are omitted throughout, since the times are very short, due to the fact that the proposed models are ``na\"ive," \textit{i.e.}, there is no learning phase. 

\subsection{Specific Location}
\label{sec:ajaccio}
In this part, only the results obtained in France at the Ajaccio site (west coast of Corsica 41°55'36"N, 8°44'13"E, alt 30 m) for different forecast horizons are shown. The climate is Mediterranean, with mild, relatively rainy winters and hot, sunny summers, sometimes sultry, but tempered by the breeze ($KG=$ Csa). The site is in an area exposed to the Mistral wind, which blows from the Gulf of Lion. The average temperature of the coldest months (January, February) is 9 °C, that of the warmest months (July, August) 22.5 °C. The forecastability $F$ is estimated close to $68\%$ \citep{doi:10.1063/5.0042710}, which gives it climatic characteristics relative to cloudy occurrences (and thus to solar radiation) relatively straightforward to predict. 

It is important to verify the impact of the ratio to seasonal trend. According to Eq.(\ref{eq:24}) and considering a $90\%$ confidence level ($\alpha=0.1$), the quantile $q_{0.95}=1.645$. Computing $\rho$ with 100 data points ($n=100$) we obtain for global solar irradiation ($I_m$) $t(m)=0.4032$ and $\rho(m)=0.62$ while in the clear sky index case ($\kappa_m$), $t(m)=0.2918$  and $\rho(m)=0.1842$. The use of clear-sky series has a significant impact on seasonality, and we therefore consider that the predictive methodology described above can be applied, as soon as the ratio to trend ($I_{CS}$) is carried out.

Tables \ref{tab:1a} and \ref{tab:1b} show the error metrics related to prediction concerning horizon between 1 and 10 h. The presented models are persistence (PER in Section~\ref{sec:subsec1}), climatology (CLIM in Section~\ref{sec:subsec2}), climatology persistence (CLIPER or AR(1) in Section~\ref{sec:AR}), exponential smoothing (ES in Section~\ref{sec:ETS}), our proposed modified ARTU (or AR(2) in Section~\ref{sec:method}) and a combination of  CLIPER, ARTU (for $R=0.05$) PER and ES (COMB in Section~\ref{sec:comb}). The MASE is computed with $m=13$ and not 24 because the filtration of night hours reduces the number of data per day, depending on the day considered (time of year); the periodicity varies between 9 and 15 h on this site.

\begin{table}[tb]
\begin{tabular}{@{}llcccccccccc@{}}\toprule
         &           & \multicolumn{10}{c}{Horizons (h)}                                                                                                                    \\  
         &           & 1             & 2             & 3             & 4             & 5             & 6             & 7             & 8             & 9             & 10            \\ \midrule
PER     & nRMSE     & 22.1          & 31.8          & 40.2          & 47.6          & 52.7          & 55.0          & 55.2          & 53.6          & 51.7          & 49.3          \\  
         & nMAE      & \textbf{11.9} & 17.6          & 21.7          & 25.1          & 27.6          & 29.3          & 30.1          & 30.3          & 29.9          & 29.2          \\ \addlinespace 
CLIM     & nRMSE     & 70.6          & 70.6          & 70.7          & 70.7          & 70.7          & 70.8          & 70.8          & 70.8          & 70.8          & 70.8          \\  
         & nMAE      & 60.7          & 60.8          & 60.8          & 60.8          & 60.9          & 60.9          & 60.9          & 61.0          & 61.0          & 61.0          \\ \addlinespace
CLIPER & nRMSE     & 21.3          & 27.7          & 30.5          & 32.5          & 33.6          & 34.0          & 34.3          & 34.5          & 34.7          & 34.8          \\  
         & nMAE      & 13.9          & 18.8          & 20.7          & 21.9          & 22.6          & 23.1          & 23.4          & 23.7          & 23.8          & 23.9          \\ \addlinespace 
ES       & nRMSE     & 22.4          & 30.3          & 33.5          & 35.5          & 36.9          & 37.0          & 36.6          & \textbf{34.4} & 36.6          & 36.9          \\  
         & nMAE      & 13.3          & 19.3          & 21.5          & 23.3          & 24.1          & 24.4          & 24.5          & 24.6          & 25.0          & 25.0          \\ \addlinespace 
ARTU     & nRMSE     & 22.2          & 30.8          & 30.7          & 32.4          & 33.5          & 34.0          & 34.3          & 34.4          & 34.7          & 34.8          \\  
R=0     & nMAE      & 14.8          & 21.3          & 20.9          & 27.8          & 22.7          & 23.2          & 23.5          & 23.6          & 23.8          & 23.9          \\ \addlinespace 
ARTU     & nRMSE     & 21.3          & 27.6          & 30.3          & \textbf{32.3} & \textbf{33.4} & \textbf{33.9} & 34.2          & 34.4          & \textbf{34.7} & 34.8          \\  
R=0.01  & nMAE      & 13.9          & 18.7          & 20.4          & 21.7          & 22.5          & 23.0          & \textbf{23.3} & \textbf{23.6} & \textbf{23.8} & \textbf{23.8} \\ \addlinespace 
ARTU     & nRMSE     & 21.3          & 27.5          & \textbf{30.3} & 32.3          & 33.4          & 33.9          & \textbf{34.2} & 34.5          & 34.7          & \textbf{34.8} \\  
R=0.05  & nMAE      & 13.9          & 18.6          & 20.4          & 21.7          & 22.5          & 23.0          & 23.3          & 23.6          & 23.8          & 23.8          \\ \addlinespace 
ARTU     & nRMSE     & 21.3          & 27.5          & 30.3          & 32.3          & 33.4          & 33.9          & 34.2          & 34.5          & 34.7          & 34.8          \\  
R=0.1   & nMAE      & 13.9          & 18.6          & 20.4          & 21.7          & \textbf{22.5} & \textbf{23.0} & 23.3          & 23.7          & 23.8          & 23.8          \\ \addlinespace 
COMB     & nRMSE     & \textbf{20.5} & \textbf{26.9} & 30.8          & 33.9          & 35.9          & 36.5          & 36.5          & 36.2          & 36.3          & 36.3          \\  
         & nMAE      & 12.4          & \textbf{17.2} & \textbf{19.5} & \textbf{21.3} & 22.6          & 23.2          & 23.6          & 23.8          & 24.0          & 24.0          \\ \bottomrule
\end{tabular}
\caption{nRMSE and nMAE for the six different benchmark methods at Ajaccio, France. The lowest error metrics values for each horizon are highlighted in bold.}\label{tab:1a}

\end{table}

\begin{table}[tb]
\centering

\begin{tabular}{@{}llccccccccc@{}}
\toprule
& PER  & CLIM   & CLIPER & ES    & \multicolumn{1}{c}{\begin{tabular}[c]{@{}c@{}}ARTU\\ R=0\end{tabular}} & \multicolumn{1}{c}{\begin{tabular}[c]{@{}c@{}}ARTU\\ R=0.01\end{tabular}} & \multicolumn{1}{c}{\begin{tabular}[c]{@{}c@{}}ARTU\\ R=0.05\end{tabular}} & \multicolumn{1}{c}{\begin{tabular}[c]{@{}c@{}}ARTU\\ R=0.1\end{tabular}} & COMB           &  \\ \midrule

MASE  & 52.10 & 124.11 & 44.45    & 46.40 & 45.23                                              & 44.29                                                 & 44.27                                                 & 44.28                                                & \textbf{43.64} &  \\ \bottomrule
\end{tabular}
\caption{MASE for the six different benchmark methods at Ajaccio, France. The lowest  value is highlighted in bold.}\label{tab:1b}

\end{table}

These tables provide a lot of information that should be confirmed by simulation of the models at other experimental sites. It is important to remember that the goal of the predictive methodologies described and tested here is not to be the best forecasting models, but the simplest ones (sometimes very simple) that would allow arbitration of the classification of more complex models. For example, forecasts made with artificial neural networks of the multilayer perceptron type on this data set are slightly less than 20\% for 1-h horizon \citep{VOYANT2018121} but are too complex to be used as reference.

\begin{figure}[tb]
\includegraphics[scale=0.45]{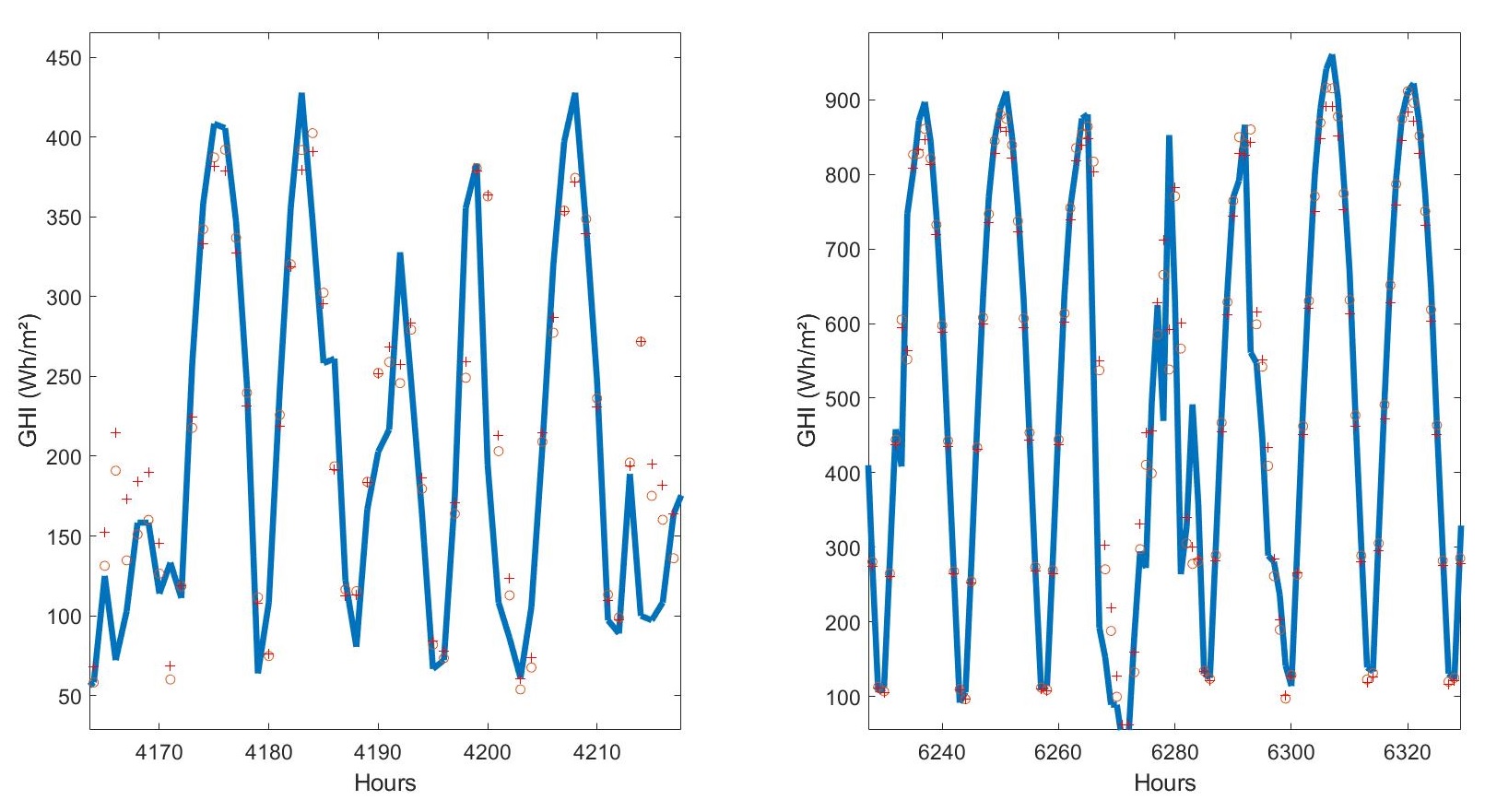}%
\caption{Prediction (1 hour horizon) for Ajaccio in winter (left) and summer (right). The blue lines correspond to the measurements (irradiation), the yellow circles to COMB, and red crosses to ARTU with $R=0.05$} 
\label{fig:fig3}
\end{figure}

First of all, climatology is not a good predictive model, but it is the simplest to implement (nRMSE $> 70\%$ whatever the horizon studied). 
At the Ajaccio site, persistence is a very good indicator for short horizons, but loses its predictive power from $h>4$. Since many energy applications (such as energy management systems of smart or micro grids) focus on horizons lower than 6 hours, this model is not excluded. 
If we want to improve the results, climatology-persistence is a very good alternative, followed closely by exponential smoothing. The ARTU model gives systematically better results than the climatology-persistence as soon as $R>0$ but in the specific case of the studied site the gain is minimal. In the following we concentrate on the case where $R=0.05$ since it is the one which proposes the lowest MASE. The combination of the models is undoubtedly the best alternative for  short horizons whereas the ARTU model is the one for larger horizons. The results of these two models is given in  Fig. \ref{fig:fig3} where a very good similarity with measurements is visible even during the most difficult winter months from a forecasting point of view. \cyr{As conclusion, within the scope of this series of simulations, the proposal to use COMB as a baseline seems appropriate considering all the horizons. }

\subsection{Multi-Site Study}
\label{sec:multi}
The results of the different forecasting methods are shown for four sites, one of which is relatively close to the previous one at about 105 km (Bastia). The other sites have entirely different climatological characteristics (Melbourne, Le Raizet and Nancy). The characteristics of these sites are shown in Table \ref{tab:tab2}.

\begin{table}[tb]
\centering
\begin{tabular}{@{}llrccc@{}} \toprule
	Name & Localisation & Coordinates & Alt (m) & $KG$ & $F$(\%) \\ \midrule
	\textbf{Ajaccio}    & France (Corsica)        & \begin{tabular}{@{}r@{}}41° 55' 36" N \\ 08° 44' 13" E  \end{tabular}  & 30                    & Csa         & 68            \\
	\textbf{Bastia}     & France (Corsica)        & \begin{tabular}{@{}r@{}}42° 39' 14" N \\ 09° 39' 59" E  \end{tabular}  & 30                    & Csa         & 62.1          \\ 
	\textbf{Nancy}      & France (Metropolitan)   & \begin{tabular}{@{}r@{}}48° 41' 31" N \\ 06° 11' 03" E  \end{tabular}  & 271                   & Cfb         & 50.2          \\ 
	\textbf{Le Raizet}  & France (Guadeloupe)     & \begin{tabular}{@{}r@{}}16° 16' 15" N \\ 61° 30' 16" W  \end{tabular}  & 11                    & Af          & 58.2          \\ 
    \textbf{Tilos}      & Greece                  & \begin{tabular}{@{}r@{}}36° 26' 00" N \\ 27° 22' 00" E  \end{tabular}  & 100                   & Csa         & 82.5          \\	
	\textbf{Melbourne}  & Australia               & \begin{tabular}{@{}r@{}}37° 48' 50" S \\ 144° 57' 47" E	\end{tabular}  & 31                    & Cfb         &  63.2         \\ \bottomrule
\end{tabular} 
\caption{Characteristics of the studied sites. Climate classification $KG$ according to \citet{ASCENCIOVASQUEZ2019672} and Forecastability $F$ according to \citet{doi:10.1063/5.0042710}}
\label{tab:tab2}
\end{table}

As can be seen from Table \ref{tab:tab3}, the conclusions stated in the previous Section remain valid. Climatology-persistence systematically improves the persistence and is itself systematically improved by the ARTU method. The combination of these methods (COMB) remains the best alternative according the results.

\begin{table}[tb]
\centering
\begin{tabular}{@{}llccccc@{}}
	\toprule
	          &       & \multirow{2}{*}{PER} & \multirow{2}{*}{CLIPER} & \multirow{2}{*}{ES} &           ARTU & \multirow{2}{*}{COMB} \\
	          &       &                       &                           &                     &     $R=0.05$ &                       \\ \midrule
	Ajaccio   &   &                 52.10 &                     44.45 &               46.40 &          44.27 &        \textbf{43.64} \\ \addlinespace
	Bastia    &   &                 67.28 &                     52.95 &               54.66 &          52.72 &        \textbf{52.36} \\ \addlinespace
	Nancy     &   &                 59.85 &                     59.25 &      \textbf{54.06} &          58.49 &                 54.23 \\ \addlinespace
	La Raizet &   &                 70.60 &                     51.76 &               55.03 & \textbf{51.52} &                 52.77 \\ \addlinespace
	Melbourne &   &                 66.12 &                     56.70 &               57.20 &          56.37 &        \textbf{54.52} \\ \bottomrule
\end{tabular}
\caption{Comparison of the 4 sites and results (MASE) for Ajaccio (see Section \ref{sec:ajaccio}). Error computed for all horizons comprise between 1 hour and 10 hours}
\label{tab:tab3}
\end{table}

\begin{figure}[htb]
\includegraphics[scale=0.5]{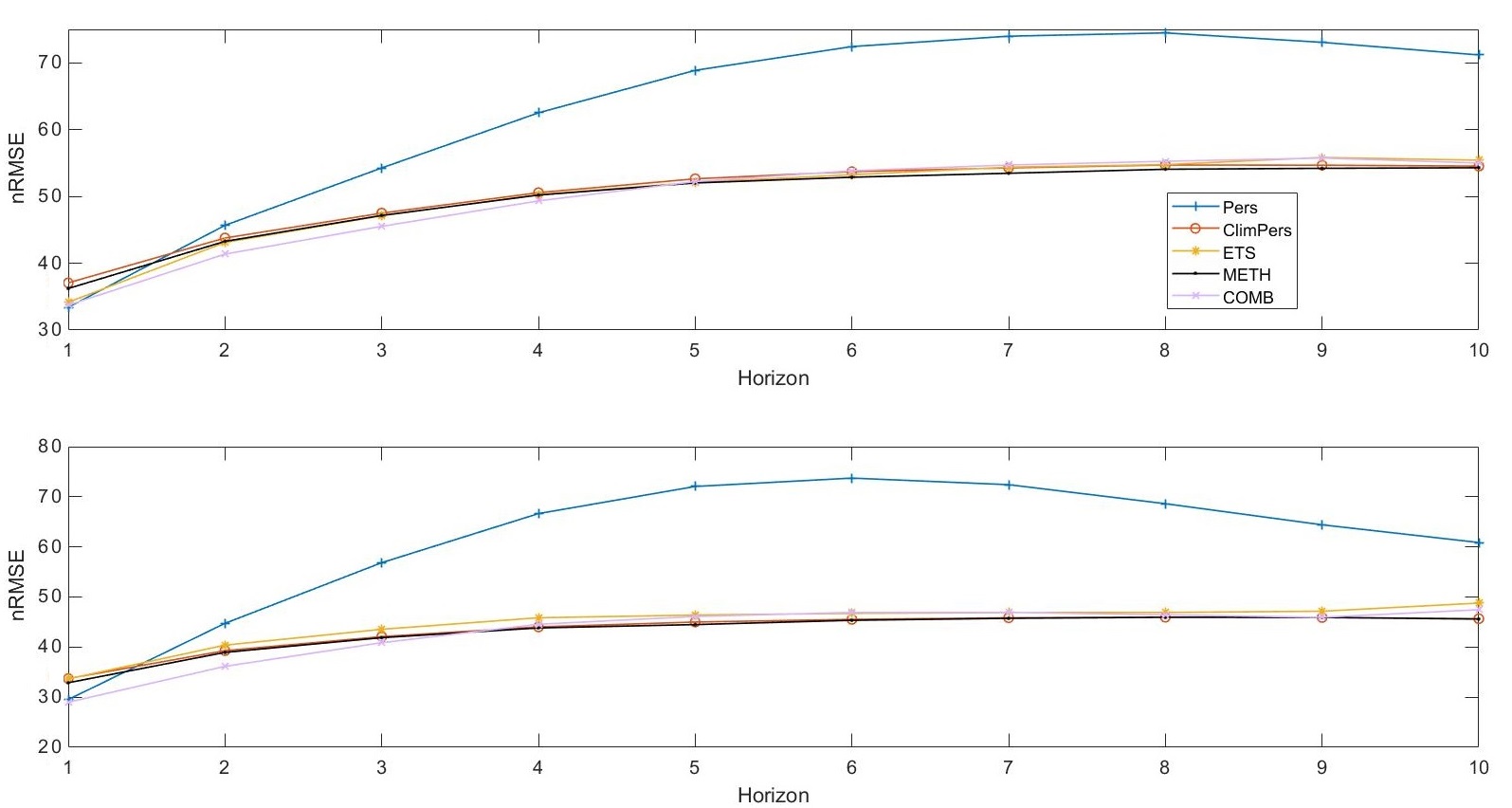}%
\caption{nRMSE \cyr{(in \%)} evolution as function of the forecast horizon for Nancy (top) and Melbourne (bottom)} 
\label{fig:fig4}
\end{figure}

If we focus on the site with the lowest forecastability (Nancy), Fig. \ref{fig:fig4} shows the evolution of the nRMSE and nMAE as  function of the forecast horizon and confirms that in this particular case the ES model is the most suitable. This surprising result is probably related to the fact that for this site the clear sky model is less efficient than for the others. Concerning the site with the lowest forecast errors (Melbourne), we see in Fig. \ref{fig:fig4} that COMB has good performance compared with the other models mainly over shorter horizons. \cyr{Drawing conclusions from this series of simulations, COMB and ARTU are the methodologies that give the best reliability.}

\subsection{Sensitivity to Change in Data Type}

In this Section, we study the sampling frequency of solar irradiance measurements. Moreover, here, we focus on inclined (or tilted) radiation (TGI of 30°) \citep{david_evaluating_2013}, which is generally more difficult to predict because the anisotropy of the sky diffusion plays important role and is difficult to quantify. No hypothesis for the nature of radiation has been formulated in Section \ref{sec:2}; it is therefore a question of checking if the preceding conclusions are valid in this particular case. The data used were measured on the Tilos site, a small Greek island with a Mediterranean climate (36°26'00"N, 27°22'00"E, alt. 100 m). This site has a forecastability $ F =  82.5$; the data were acquired every 15 minutes and the climate of the site has a $KG$ of Csa type. For the MASE calculation, $m$ (the period in Eq.(\ref{eq:26})) was modified and taken equal to 52 ($= 13$x$4$) because there are 4 measurements per hour. The results are available in Table \ref{tab:4}.

\begin{table}[tb]
	\centering
	\begin{tabular}{@{}llccccc@{}}
		\toprule
		\multirow{2}{*}{Horizon} & \multirow{2}{*}{Metrics} & \multirow{2}{*}{PER} & \multirow{2}{*}{CLIPER} & \multirow{2}{*}{ES} &      ARTU      & \multirow{2}{*}{COMB} \\
		                         &                          &                       &                           &                     &   $R=0.05$   &                       \\ \midrule
		15 min                   & nRMSE                    &    \textbf{13.52}     &           19.27           &        24.86        &     18.71      &         14.00         \\
		                         & nMAE                     &     \textbf{6.13}     &           12.16           &        13.97        &     11.75      &         8.26          \\ \addlinespace
		75 min                   & nRMSE                    &         58.87         &           24.99           &        38.36        & \textbf{24.89} &         26.98         \\
		                         & nMAE                     &         20.91         &           15.83           &        23.15        &     15.73      &    \textbf{15.34}     \\ \addlinespace
		150 min                  & nRMSE                    &         116.9         &           25.98           &        37.55        & \textbf{25.95} &         40.17         \\
		                         & nMAE                     &         36.00         &           16.39           &        23.71        & \textbf{16.38} &         19.91         \\ \addlinespace
		                         & MASE                     &         48.49         &           33.04           &        47.52        & \textbf{32.61} &         33.04         \\ \bottomrule
	\end{tabular}
    \caption{Error metrics from the Tilos evaluation. Global tilted irradiance measured every 15 minutes.}
    \label{tab:4}
\end{table}

In this specific case, the fact that the measurements are so close together dramatically changes the results. Persistence is by far the best model for the 1-h horizon but it becomes the worst as the horizon increases. The best compromise seems to be COMB, although relatively bad at 150 min. In operations, it is not uncommon to study time steps of 15 minutes, particularly for piloting solar energy station (power or energy management system) but not up to a horizon of 150 min. In this case it will be necessary to use hourly data. Based on this observation, the COMB model is the one that provides more reliability.

\subsection{Concerning Other Meteorological Data}
\label{sec:other}
This section is dedicated to the study of meteorological time series other than solar. These are hourly series of ambient temperature \citep{CORCHADO1999351} and 10-meter wind speeds \citep{Giebel_2016}. Temperature is largely periodic and similar what we observed with solar radiation, though wind speed is entirely different and exhibits little difference between day and night. The chosen site is Nancy, already studied in Section \ref{sec:multi}. The MASE evaluation is calculated in this case with its non-periodic version. Here we quantify the importance of the stationarity of the data. Table \ref{tab:5} presents the temperature results related to the different models assuming $I_{CS}=1$ ($\forall t$). This corresponds to the non-stationary case in which $x_t$ and $\kappa_t$ are considered equal. We can expect that all models are penalized from the absence of deseasonalization of the time series, which particularly impacts ARTU.

\begin{table}[tb]
    \centering
    \begin{tabular}{@{}llccccc@{}}
        \toprule
        \multirow{2}{*}{Horizon} & \multirow{2}{*}{Metrics} & \multirow{2}{*}{PER} & \multirow{2}{*}{CLIPER} & \multirow{2}{*}{ES} &      ARTU      & \multirow{2}{*}{COMB} \\
                                 &                          &                       &                           &                     &   $R=0.05$   &                       \\ \midrule
        15 min                   & nRMSE                    &         8.90          &           8.85            &        9.03         &     17.16      &     \textbf{8.79}     \\ 
                                 & nMAE                     &         6.32          &       \textbf{6.26}       &        6.45         &     13.59      &         6.35          \\ \addlinespace
        75 min                   & nRMSE                    &         27.91         &           26.21           &        27.16        &     26.25      &    \textbf{26.13}     \\ 
                                 & nMAE                     &         21.90         &           20.36           &        21.60        & \textbf{20.07} &         20.30         \\ \addlinespace
        150 min                  & nRMSE                    &         39.28         &           34.36           &   \textbf{29.36}    &     30.67      &         31.45         \\ 
                                 & nMAE                     &         33.62         &           28.17           &        25.09        & \textbf{24.80} &         26.39         \\ \addlinespace
                                 & MASE                     &         352.4         &           310.9           &        312.5        & \textbf{290.9} &         307.2         \\ \bottomrule
    \end{tabular}
    \caption{Error metrics from the temperature evaluation without seasonal adjustments.}
    \label{tab:5}
\end{table}

 The results do not correspond to what we have observed so far. The dynamic range of the signal tested is very low and the regularity of the successive measurements make it a good candidate for the use of persistence. Although the results are not good for the very short-term horizons; ARTU remains the method which is the most efficient on average. However, given its simplicity and the very good results observed in the first horizons, ES is undoubtedly the reference method that should be used to properly characterize temperature forecasts. This suggests that as soon as the signal is regular with low variability, the forecast reference must imperatively be made with exponential smoothing. If we now focus on the estimation of wind speeds (Table \ref{tab:6} still considering $I_{CS}=1$), the conclusions are different yet not inconsistent given the large difference in variability between these two meteorological quantities. 

\begin{table}[tb]
\centering
    \begin{tabular}{@{}llccccc@{}}
        \toprule
        \multirow{2}{*}{Horizon} & \multirow{2}{*}{Metrics} & \multirow{2}{*}{PER} & \multirow{2}{*}{CLIPER} & \multirow{2}{*}{ES} &      ARTU      & \multirow{2}{*}{COMB} \\
                                 &                          &                       &                           &                     &   $R=0.05$   &                       \\ \midrule
        15 min                   & nRMSE                    &         46.47         &      \textbf{41.78}       &        44.28        &     47.00      &         42.33         \\ 
                                 & nMAE                     &         31.98         &      \textbf{30.18}       &        32.19        &     34.62      &         30.47         \\ \addlinespace
        75 min                   & nRMSE                    &         70.23         &           52.67           &        53.59        & \textbf{52.65} &         54.22         \\ 
                                 & nMAE                     &         53.22         &           39.42           &        40.12        & \textbf{39.39} &         40.85         \\ \addlinespace
        150 min                  & nRMSE                    &         73.73         &      \textbf{53.04}       &        53.96        &     53.16      &         55.10         \\ 
                                 & nMAE                     &         56.31         &      \textbf{39.53}       &        40.49        &     39.58      &         41.50         \\ \addlinespace
                                 & MASE                     &         159.6         &           120.2           &        122.9        & \textbf{120.2} &         124.5         \\ \bottomrule
    \end{tabular}
\caption{Error metrics concerning the wind speed evaluation without seasonal adjustments.}
\label{tab:6}
\end{table}

We observed that the climatology-persistence performed well in the case of solar radiation and its prediction. We see here that this is the best of the tested models and undoubtedly the easiest to set up. The ARTU model gives fairly comparable results. If, in the case of temperature, the ACF values were close from one lag to another, here it is the reverse; the correlations become insignificantly different from 0 very quickly. We observe an exponential decay (see Section \ref{sec:method}), which suggests that the best model is indeed an AR(1) and therefore it is not surprising that CLIPER performs best and that ARTU does not bring a real added value to this case.  

Using $I_{CS}$ as defined in Section \ref{sec:season} with the time decomposition, we are able to see the impact of seasonal adjustment on the results. Tables \ref{tab:7} and \ref{tab:8} show the results according the temperature and the wind speed, respectively.

\begin{table}[tb]
\centering
    \begin{tabular}{@{}llccccc@{}}
    	\toprule
    	\multirow{2}{*}{Horizon} & \multirow{2}{*}{Metrics} & \multirow{2}{*}{PER} & \multirow{2}{*}{CLIPER} & \multirow{2}{*}{ES} &      ARTU      & \multirow{2}{*}{COMB} \\
    	                         &                          &                       &                           &                     &   $R=0.05$   &                       \\ \midrule	
    	15 min                   & nRMSE                    &         6.08          &       \textbf{5.91}       &        6.19         &      6.85      &         6.02          \\ 
    	                         & nMAE                     &     \textbf{4.01}     &           4.03            &        4.17         &      5.02      &         4.06          \\ \addlinespace
    	75 min                   & nRMSE                    &         14.66         &           12.19           &        13.86        & \textbf{12.17} &         12.82         \\ 
    	                         & nMAE                     &         10.25         &           9.06            &        9.82         & \textbf{9.06}  &         9.23          \\ \addlinespace
    	150 min                  & nRMSE                    &         19.62         &           14.07           &        14.66        & \textbf{13.12} &         14.44         \\ 
    	                         & nMAE                     &         14.01         &           10.74           &        11.01        & \textbf{9.96}  &         10.87         \\ \addlinespace
    	                         & MASE                     &         149.5         &           131.8           &        134.5        & \textbf{129.3} &         130.8         \\ \bottomrule
    \end{tabular}
\caption{Error metrics concerning the temperature evaluation with seasonal adjustments.}
\label{tab:7}
\end{table}

Even though this result has been understood for some time, we measure the importance of the ratio to trend and of the seasonal adjustment with these 2 Tables. The temperature results are very good for all methods and particularly for ARTU (nMAE$=9.96$ for a 10-h horizon). It is likely that few machine-learning methods can significantly improve this result. For wind speeds, CLIPER remains the best way to make simple forecasts, even if ARTU and COMB can be very interesting alternatives, but also more complicated to set up. 
\cyr{As in the previous sub-sections, a conclusion to propose COMB and ARTU as the best compromise of reference methods can be made based on the results of this section.}

\begin{table} [ht]
\centering
    \begin{tabular}{@{}llccccc@{}}
    	\toprule
    	\multirow{2}{*}{Horizon} & \multirow{2}{*}{Metrics} & \multirow{2}{*}{PER} & \multirow{2}{*}{CLIPER} & \multirow{2}{*}{ES} &      ARTU      & \multirow{2}{*}{COMB} \\
    	                         &                          &                       &                           &                     &   $R=0.05$   &                       \\ \midrule	
    	15 min                   & nRMSE                    &         39.87         &           35.11           &        36.30        &     34.95      &    \textbf{34.90}     \\ 
    	                         & nMAE                     &         28.64         &           25.41           &        26.14        &     25.30      &    \textbf{25.18}     \\ \addlinespace
    	75 min                   & nRMSE                    &         55.21         &      \textbf{41.32}       &        42.24        &     41.33      &         42.10         \\ 
    	                         & nMAE                     &         39.91         &           29.70           &        30.59        & \textbf{29.69} &         30.51         \\ \addlinespace
    	150 min                  & nRMSE                    &         58.53         &      \textbf{42.07}       &        42.85        &     42.26      &         43.31         \\ 
    	                         & nMAE                     &         42.79         &      \textbf{30.09}       &        30.84        &     30.21      &         31.30         \\ \addlinespace
    	                         & MASE                     &         118.8         &           90.0            &        91.4         & \textbf{89.7}  &         91.7          \\ \bottomrule
    \end{tabular}
\caption{Error metrics concerning the wind speed evaluation with seasonal adjustments.}
\label{tab:8}
\end{table}

The field of wind forecasting is a very particular and complicated discipline. The choice to use the mean to try to make the series stationary is simple to establish though is perhaps not the best approach should we seek better results. Though not mentioned here, single, double or seasonal differencing would all have the effect of improving stationarity.

\section{Conclusions}
\label{sec:conclusion}

In this paper, we proposed a new way of evaluating forecasts made in the field of meteorology and more specifically on solar radiation (Statistical Reference Method i.e. SRM). As the current practice is to compare the results of elaborate forecasting methods with those of simple methods (naive or reference), it should be appreciated that simple methods can evolve and improve. In the solar world for example, the default reference model is to use persistence applied to the clear-sky index (sometimes also called smart persistence or scaled persistence). Recently, a new reference emerged from \cite{YANG2019981} that updated what was initially proposed by \cite{murphy_climatology_1992}. 

In this paper, we evaluated all reference methods (PER, CLIM, CLIPER and ES) under the same formalism based on putting forward the measurement error, seasonality and forecast error through covariance estimation. This resulted in an evolution of the technique and the proposition of a new, simple model that mixes prediction and filtration, which in the end is quite similar to an AR(2) with the additional quality that no learning process is necessary. We title this new method ARTU which is the main innovation of this work.

The use of the classical exponential smoothing (ES) tools has been demonstrated to be of use again. Though it has been used in some renewable energy papers \citep{DONG20131104}, it is not included as part of the references nor of the naive methods. However, we have shown in this paper that ES belongs in that category.

The choice to consider combination (COMB) as a  \cyr{benchmark} method can be seen as a ``curiosity'', but in conclusion, when we combine two (or more) benchmarks, \cyr{the result}   \cyr{meets the criteria to be considered a benchmark method}. Averaging makes it possible to overcome outliers and smooth forecasts. After this study, we can recommend the use of the combination of simple models such as PER, CLIPER, ES and ARTU.  \cyr{The main rules and recommendations of this study are}:
\begin{itemize}[label=$\looparrowright$]  
\item Reference models are critical to justify any new forecasting approach;
\item The most appropriate benchmarking method depends on the nature of the variable being forecasted (seasonality, periodicity, forecastability) as well as the forecast horizon;
\item  The use of exponential smoothing for the estimation of ``regular" series (high forecastability) should not be overlooked;
\item The use of one error metrics specific to the time series forecasting community (MASE ) are real assets deserving of uptake in more applied fields like solar energy engineering;
\item The forecaster should endeavor to discover the benchmarking method most appropriate for their needs by either using them all (PER, CLIPER, ES, ARTU and COMB) and taking the best approach or by justifying the selection;
\item \cyr{In conclusion, COMB appears to be the best method among all those tested.}
\end{itemize}

The development of increasingly sophisticated forecasting methods is necessary, but it is crucial to be able to evaluate them and to benchmark their efforts against the expected results. It is important to offer the most efficient benchmark in order to be able to compare studies with each other and identify the most efficient models even if it comes down to a fine detail. It is important to mention that in the solar energy community, it is usual to consider that an improvement of 1\% in the prediction is approximately equivalent to an increase of 2\% in economical gain \citep{DAVID2021672}. For very large installations the gain can quickly become substantial.

\bibliography{bib}


 \appendix
 
 \section{Proof of CLIPER}
 \label{appendix:cliper}
 From Eq.(\ref{eq:eq4bis}):
 \begin{equation}
\mathbb{E}[x^c_{t+h}x^c_{t}]=\mathbb{E}[\alpha x^c_tx^c_t+\omega_{t+h}x^c_t],
\end{equation}
 and from the definition of a white noise ($\omega$ uncorrelated with other variables) and of a stationary process ($x^c$), it becomes $\mathbb{E}[\omega_{t+h}x^c_t]=0$. Then let $\sigma_{x}^2$ and $\gamma(h)$ be, respectively, the variance of $x_t$ (equal to the variance of $x^c_t$) and the autocovariance factors \citep{MakridakisSpyrosG1998F} between $x^c_t$ and $x^c_{t+h}$ \footnote{or $x^c_{t-h}$ because autocorrelation is an even function while $x^c_t\in \mathbb{R}$ is a wide-sense stationary process}, we observe that  $\mathbb{E}[x^c_{t}x^c_{t}]=\sigma_x^2$ and as $x^c_t$ is a centered variable, that $\mathbb{E}[x^c_{t}x^c_{t+h}]=\gamma(x^c_{t},x^c_{t+h})+(\mathbb{E}[x^c_t])^2\equiv\gamma(h)$.
Hence, Eq.(\ref{eq:eq4bis}) can be modified and simplified as ($\mathbb{E}[z_{t+h}]=\mathbb{E}[z_t]$):

\begin{equation}
\label{eq:eq4ter}
\gamma(h)=\mathbb{E}[\alpha x^c_tx^c_t]+\mathbb{E}[\omega_{t+h}x^c_t]=\alpha \sigma_x^2. 
\end{equation}

This implies that there is a simple link between $\alpha$ and the autocorrelation factor $\rho$ (defined from the ratio between the autocovariance and the variance):

\begin{equation}
\label{eq:eq4quad}
\alpha=\gamma(h)/\sigma_x^2=\rho(h).
\end{equation}

Notice that considering the AR(1) model in Eq.(\ref{eq:eq1}) and the Yule--Walker equations or using Eq.(\ref{eq:eq4quad}) one can conclude that the prediction is now related to: 
\begin{equation}
\label{eq:eq44}
\widehat{x}^c_{t+h}=\rho(h)  x^c_t,
\end{equation}
\noindent as $x^c_t=x_t-\mathbb{E}[x]$ and $\mathbb{E}[x]=\mathbb{E}[y]$, Eq.(\ref{eq:eq44}) can be replaced with:

\begin{equation}
\label{eq:eq45}
\widehat{x}_{t+h}=\rho(h)  \left[x_t-\mathbb{E}(x)\right] +\mathbb{E}(x)=\rho(h)  \left[x_t-\mathbb{E}(y)\right] +\mathbb{E}(y).
\end{equation}

Let's not forget that $x_t$ is an unknown quantity but that it is possible to have a measure of it through the variable called $y_t$ (see Eq.(\ref{eq:eq2})), this leads us to the following result:

\begin{equation}
\label{eq:eq4}
\widehat{x}_{t+h}=\rho(h)y_t+\left[1-\rho(h)\right]\mathbb{E}(y).
\end{equation}

From a classical point of view, this equation is no different than the forecast by an AR(1) model using Yule--Walker equations, in the case of solar radiation prediction, this predictor (convex combination\footnote{In convex geometry, a convex combination is a linear combination of points in an affine space where all coefficients are non-negative and sum to 1} between climatology and persistence) is a particular reference called ``climatology--persistence combination". It tends to become the reference model in comparative studies \citep{YANG2019981}.

Note that by taking the expectation of Eq.(\ref{eq:eq1}) multiplied by $x^c_{t+h}$ and because  $\mathbb{E}[\omega_t x^c_t]=\sigma^2_{\omega}$, it becomes:

\begin{equation}
\label{eq:eq5}
\sigma_x^2=\alpha \gamma(h)+\sigma^2_{\omega}.
\end{equation}

Replacing $\alpha$ with $\rho(h)$ in Eq.(\ref{eq:eq5}), we obtain: 

\begin{equation}
\label{eq:eq6}
\sigma^2_{\omega}=\sigma_x^2[1-\rho(h)^2],
\end{equation}

\noindent where $\sigma^2_{\omega}$ corresponds to the forecast error estimated by the Euclidean norm (mean square error; MSE) in the unbiased case.
Because the correlation coefficient $\rho(h)$ is between 0 and 1, $\sigma^2_{\omega}$ can take values between $\sigma_x^2$ ($\widehat{x}_{t+h}=\bar{y}$, see Section \ref{sec:subsec2}) and 0 ($\widehat{x}_{t+h}=y_t$, see Section \ref{sec:subsec1}). For example (see \ref{algo:ClimPers} for details), when $\rho(h)$ takes the value of 0.7, the model is written $\widehat{x}_{t+h}=0.7y_t+0.3\mathbb{E}(y)$ and the expected error is about $0.5\sigma_x^2$. Of course, this theoretical error is not observed in reality, since we make the assumption that the observed phenomenon can be modeled by an AR(1), which is not necessarily true and which thus induces an additional part of uncertainty and an increase in the observed error. 
Moreover, the measurement error, represented by the parameter $\sigma^2_{v}$ in Eq.(\ref{eq:eq2}), also penalizes the forecast error.
 
 \section{Proof of ARTU}
 \label{proof}
Equations \ref{eq:minus}-\ref{eq:eq8c} allow one to understand the link that can exist between a classical AR(2) and the method that we propose in this section: the prediction ($\widehat{x}_{t+h}$) depends on two previous measures ($y_t$ and $y_{t-h}$). The goal of this method is to find a mathematical formulation for $\alpha$ and $K$. A classic way to do this is to minimize the mean square error (MSE) which takes the form:

\begin{equation}
\label{eq:eq9}
\textnormal{MSE}=\mathbb{E}\left[(x^c_{t+h}-\widehat{x}^c_{t+h})^2\right]=\mathbb{E}\left[(x^c_{t+h}-\alpha x^c_t-K(y^c_t-\widehat{x}^{c -}_{t}))^2\right].
\end{equation}

From Eq.(\ref{eq:eq2}), considering $v_t$ (variance $\sigma^2_{v}$) has a covariance with all random variables null, Eq.(\ref{eq:eq9}) can be replaced with Eq.(\ref{eq:eq10}). This simplification is possible under the assumption of orthogonality (or uncorrelation working with centered series). 

\begin{equation}
\label{eq:eq10}
\textnormal{MSE}=K^2\sigma^2_{v}+\mathbb{E}(u)^2 ~\textnormal{with} ~u=x^c_{t+h}-\alpha x^c_t-x^cz_t+K\alpha x^c_{t-h}.
\end{equation}

A sufficient condition to find the optimal values of $\alpha$ and $K$ consists of finding their values that minimize the MSE, by equating to 0 its gradient. This is equivalent to solving $\partial \textnormal{MSE}/\partial K=0$ (see Eq.~\ref{eq:eq11}) and $\partial \textnormal{MSE}/\partial \alpha=0$ (see Eq.~\ref{eq:eq12}). Note that interchanging the derivative with expectation can be done using the dominated convergence theorem\footnote{It is one of the main theorems of Lebesgue's integration theory giving a sufficient condition for the convergence of expected values of random variables.}.

\begin{equation}
\label{eq:eq11}
\frac{\partial \textnormal{MSE}}{\partial K}=2K\sigma^2_{v}+\frac{\partial \mathbb{E}(u)^2}{\partial K}=2K\sigma^2_{v}+\mathbb{E}\left(\frac{\partial u^2}{\partial u}\frac{\partial u}{\partial K}\right)=2K\sigma^2_{v}+2\mathbb{E}[u(\alpha x^c_{t-h}-x^c_t)]=0.
\end{equation}

An identical reasoning allows us to establish Eq.(\ref{eq:eq11}) and Eq.(\ref{eq:eq12}) concerning the derivative with respect to $K$ and $\alpha$. 

\begin{equation}
\label{eq:eq12}
\frac{\partial \textnormal{MSE}}{\partial \alpha}=2\mathbb{E}[u(K x^c_{t-h}-x^c_t)]=0.
\end{equation}

After some mathematical simplifications, the solution of the problem amounts to finding $\alpha$ and $K$ solutions of the system described by Eq.(\ref{eq:eq13}). It is a system of quadratic equations with 2 unknowns ($K$ and $\alpha$) of degree 2.

\begin{equation}
\label{eq:eq13}
\left \{
\begin{array}{rcl}
K(\sigma^2_{v}+\sigma_x^2)+\alpha\big(\gamma(2h)+\sigma_x^2\big)-2K\alpha\gamma(h)-\alpha^2\gamma(h)+K\alpha^2\sigma_x^2&=&\gamma(h) \\
K\big(\gamma(2h)+\sigma_x^2\big)-2K\alpha\gamma(h)+\alpha\sigma_x^2-K^2\gamma(h)+K^2\alpha\sigma_x^2&=&\gamma(h)
\end{array}
 \right.
\end{equation}

Solving this system is not trivial, and it is best to slightly modify it to make the task easier. The more convenient form is certainly obtained dividing the two equations by $\sigma_x^2$ ($\in \mathbb{R^*}$) and defining a new variable $R=\sigma^2_{v}/\sigma_x^2$. A discussion about it is proposed in the Section \ref{sec:method}. This modification allows to highlight the auto-correlation coefficients $\rho$ as shown in Eq.(\ref{eq:eq14}). $\rho(h)$ is related to the correlation between $x^c_t$ and itself delayed by $h$ lags, while $\rho(2h)$ concerns a delay of $2h$ lags.

\begin{equation}
\label{eq:eq14}
\left \{
\begin{array}{rcl}
K(1+R)+\alpha[1+\rho(2h)]-2K\alpha\rho(h)-\alpha^2\rho(h)+K\alpha^2&=&\rho(h) \\
K[1+\rho(2h)]-2K\alpha\rho(h)+\alpha-K^2\rho(h)+K^2\alpha&=&\rho(h)
\end{array}
\right.
\end{equation}

Among the five couples of solutions, we retain only those that have a physical reality: the real ones (most of the time there are $2$). Rather than finding exact and symbolic solutions since there are no simple ones (according to our knowledge and what we could find in the literature), we propose to make them explicit by means of Levenberg–Marquardt algorithm \citep{10.1007/BFb0067700}. We used $100$ iterations with random initializations of the $2$ unknowns ($K$ and $\alpha$) between $-1$ and $1$. This method requires at each iteration to compute $(\textnormal{J}^\top \textnormal{J}+\lambda \textnormal{diag}(\textnormal{J}^\top \textnormal{J})^{-1})\textnormal{J}^\top$ where the Jacobian (J) is defined in Eq.(\ref{eq:eq14}) such as:  

\begin{equation}
\label{eq:eqJ}
\textnormal{J}=
 \begin{pmatrix} \alpha^2-2\rho(h)\alpha+1+R & \rho(2h) + 2K\alpha - 2K\rho(h) - 2\alpha\rho(h) + 1 \\ \rho(2h) + 2K\alpha - 2K\rho(h) - 2\alpha\rho(h) + 1 & K^2-2K\rho(h)+1. 
 \end{pmatrix}.
\end{equation}

The Monge theorem allows to retain only the critical points compatible with a local minimum of the MSE. In mathematical analysis, this theorem is used to study the behavior of a function with two variables ($K$,$\alpha$) in the neighborhood of a critical point ($K^*,\alpha ^*$). The retained solutions satisfy the condition of the local extremity mentioned in Eq.(\ref{eq:monge}) and of strict local minimum point exposed in Eq.(\ref{eq:monge2}).

\begin{equation}
\label{eq:monge}
\left[\frac{\partial^2 \textnormal{MSE}}{\partial K \partial \alpha}(K^*,\alpha^*)\right]^2-\frac{\partial^2 \textnormal{MSE}}{\partial K^2 }(K^*,\alpha^*)\frac{\partial^2 \textnormal{MSE}}{\partial \alpha^2 }(K^*,\alpha^*)<0,
\end{equation}

\begin{equation}
\label{eq:monge2}
\frac{\partial^2 \textnormal{MSE}}{\partial K^2 }(K^*,\alpha^*)>0.
\end{equation}

In case there is more than one solution (according to a particular triplet [$R$,$\rho(h)$,$\rho(2h)$]), we choose to calculate the associated MSE and to retain only those which minimize this quantity. This can be done from the two partial derivatives defined in Eq.(\ref{eq:eq14}) and respectively in Eq.(\ref{eq:partial1}) and Eq.(\ref{eq:partial2}) concerning $K$  and $\alpha$.

\begin{equation}
\label{eq:partial1}
\frac{\partial \textnormal{MSE}}{\partial K}=K(1+R)+\alpha[1+\rho(2h)]-2K\alpha\rho(h)-\alpha^2\rho(h)+K\alpha^2-\rho(h),
\end{equation}

\begin{equation}
\label{eq:partial2}
\frac{\partial \textnormal{MSE}}{\partial \alpha}=K[1+\rho(2h)]-2K\alpha\rho(h)+\alpha-K^2\rho(h)+K^2\alpha-\rho(h).
\end{equation}

A succession of integration (with respect to $K$ then with respect to $\alpha$) allows to determinate without too much difficulty a function (MSE in our case) defined by its partial derivatives. The result is given in Eq.(\ref{eq:eq16}) with a constant of integration $c$ ($\in \mathbb{R}$).

\begin{equation}
\label{eq:eq16}
\textnormal{\textnormal{MSE}}=K^2\left[\frac{R}{2}+\frac{1}{2}\right]-K\rho(h)-\alpha[K^2\rho(h)-K[\rho(2h)+1]+\rho(h)]+\alpha^2\left[\frac{K^2}{2}-K\rho(h)+\frac{1}{2}\right]+c
\end{equation}
 
 \section{Seasonality and Stationarity}
\label{sec:season}
The methods presented in Section \ref{sec:2} are based on the assumption that the time series studied are non-seasonal and devoid of trend. The latter hypothesis is always the case in meteorology on short time scales and particularly in solar radiation (stability of the Holocene climate) that has no significant inter-annual change (the annual average of the signal can be considered constant over a period of 10 years). The first hypothesis on the other hand is by nature invalid for weather series. Fortunately in solar energy forecasting, a transformation for removing the seasonality can be calculated very easily, via a clear-sky model \citep{SUN2021110087}. Clear-sky irradiance ($I_{CS}$) is the solar radiation incident on a horizontal surface under a cloud-free sky. Thus, the global horizontal irradiance ($I_{GH}$) are related to the seasonally adjusted variable, namely, the clear-sky index ($\kappa$), through the following:

\begin{equation}
\label{eq:S1}
    x\equiv \kappa =I_{GH}/I_{CS},
\end{equation}

If $I_{CS}$ is well-modeled, in theory $x_t$ (the clear sky index) and $y_t$ (its measurement) are without seasonality and are bounded between $ 0 $ and somewhere between $ 1 $ and $1.5$---the upper bound would depend on the cloud meteorology; in practice, an upper bound of 1.2 is often used \citep{doi:10.1063/1.5094494}. However, in terms of seasonality, it has been shown that even the best clear-sky models today are unable to completely remove it, resulting in a nonstationary clear-sky index time series \citep{doi:10.1063/5.0003495}. In practice, $I_{CS}$ is often calculated from the Lambert--Beer type relations \citep{ineichen_broadband_2008} or using directly the data from CAMS McClear service \citep{Lefevre2013}. 

 In the case of studying temperature and wind speed, the methodology is equivalent. $I_{CS}$ is replaced by the measure of the meteorological quantity by calculating for each hour of the year the average of the same hours concerning the previous years (model free). Generally, this procedure behaves as a particular kind of low-pass filter, and $I_{CS}$ is equivalent to smoothed series of temperature and wind speed.



The impact of seasonal adjustment can be quantified from Eq.(\ref{eq:24}) and the statistic $t(m)$ \citep{article7} or computing the squared $m$-th auto-correlation of the series and comparing it to a $\chi(1) ^2$ distribution as described in \cite{article8}.

\begin{equation}
\label{eq:24}
    t(m)=q_{1-\alpha/2}\sqrt{\frac{1+2\sum_{i=1}^{m-1}\rho^2(i)}{n}},
\end{equation}
where $q$ is the quantile function of the standard normal distribution and $100(1 - \alpha)\%$ corresponds to the confidence level; a $90\%$ confidence level is often used. $m$ is the number of the periods within a seasonal cycle (for example,  24 and $8760=24$x365 for hourly data).
So, the larger the value of $t(m)$, the larger the seasonality is. If $\lvert \rho(m)\lvert<t(m)$, the series can be considered deseasonalized. However, one must be careful because, like all statistical tests, this test is very sensitive to the size of the sample, so it is more relevant to subsample (randomly) the data if one wants a better interpretation of the test. We can assume $n=100$ without loss of generality. 

It is important to note that the forecasting of solar radiation time series is a special topic. The models presented previously and simulated in the next Section do not refer to a consideration of seasonality because of the clear-sky model. Forecasting the clear-sky index, and using the Error--Trend--Seasonal (ETS, N=none, A=additive) framework terminology of \citet{Hynd2008}, PER and CLIM are (N,N,N), CLIPER, ES, ARTU and COMB are (A,N,N). None of them require an optimization and training phases and some of them can be used with only recent measurements.

 \section{Algorithms}
 \label{appendix}
 The pseudo-codes detailed in this section are adapted to the case of global irradiation ($I_{GH}(t)$ with $t\in \{\overbrace{1,...,T^*,}^{InSample}\overbrace{T^*,...,T}^{OutSample}\}$) though they can be modified for any kind of time series. Up to now, we deliberately neglect the phenomena of over irradiance ($I_{GH}(t) \in [0,I_{CS}(t)]$), however depending on the time step, the clear sky model used and the quality of the time-stamp, it could be necessary to multiply $ I_{CS} $ by an arbitrary coefficient $\beta$ (generally between 1 and 2). Anyway, all models must make it possible to provide forecasts for all hours of the day and night, however the validation of the results is only performed on the daytime hours (authorizing solar elevation greater than $5-10^\circ$). Even if it is not the purpose of this study, it is important not to neglect the forecasts of the first and last hours of the daylight, they can be very important for energy management systems. Often, the real reasons for which a filtration is operated because of the poor quality of the detection concerning these hours and the strong repercussions (periodic peaks on $\kappa$) that a poor time-stamp can induce.

\subsection{Persistence}
\label{algo:Pers}
This persistence predictor (Algorithm \ref{algo1}) is certainly the simplest method use in order to operate predictions with reliability. 

\begin{algorithm}
\caption{PER}
\begin{algorithmic} 
\label{algo1}
\REQUIRE $ I_{CS}, I_{GH},h>0, \beta \in [1,2] $
\ENSURE $\widehat{I}_{GH}(t+h)$ with $t \in [T^*,T]$
\STATE {$n \leftarrow 0$}
\REPEAT 
\STATE $n\leftarrow n+1$
\UNTIL {$I_{CS}(t-n) \ne 0$}
\STATE {Pred$ \leftarrow min(I_{GH}(t-n)\times I_{CS}(t+h)/I_{CS}(t-n),\beta \times I_{CS}(t+h))$}
\RETURN {Pred}
\end{algorithmic}
\end{algorithm}

\subsection{Climatology}
\label{algo:Clim}
Even if this predictor (Algorithm \ref{algo2}) is never used in practice, it is an important way to gauge results in solar prediction study. When no model of knowledge is available, a moving average can be a good alternative. One of the characteristics of this model is that the observed forecast error is constant whatever the horizon considered. The filtering parameter ($\epsilon$) is usually taken close to 10 (Wh/m$^2$) while certain authors prefer use a threshold between $5^\circ$ and $10^\circ$ concerning the solar elevation. The information linked to the cloudiness being observable only in the presence of daylight, only these moments must be used. This means that at sun-up, it is the data from the day before that is used, so we understand the limit of statistical forecast models using only endogenous quantities.

\begin{algorithm}
\caption{CLIM}
\begin{algorithmic} 
\label{algo2}
\REQUIRE $ I_{CS},I_{GH},h>0,\epsilon \in [1,30]  $
\ENSURE $\widehat{I}_{GH}(t+h)$ with $t \in [T^*,T]$
\FOR{$n:=1$ to $T^* $}
\IF{$I_{CS}(n)<\epsilon$}
\STATE{$\kappa(n)=\emptyset$}
\ELSE
\STATE $\kappa(n) \leftarrow I_{GH}(n)/I_{CS}(n)$
\ENDIF
\ENDFOR
\STATE{$\bar{\kappa} \leftarrow $mean($\kappa(n))$}
\STATE Pred$\kappa \leftarrow \bar{\kappa}$
\STATE {Pred $\leftarrow$ Pred$\kappa \times I_{CS}(t+h)$}
\RETURN {Pred}
\end{algorithmic}
\end{algorithm}

  \subsection{Climatology Persistence}
 \label{algo:ClimPers}
CLIPER is undoubtedly the new standard of reference forecast for solar irradiation. As in the previous case (\ref{algo:Clim}), a filtering process is operated in Algorithm \ref{algo3} and  the $\epsilon$ parameter is considered for this task.
 
\begin{algorithm}
\caption{CLIPER}
\begin{algorithmic} 
\label{algo3}
\REQUIRE $ I_{CS}, I_{GH},h>0, \beta \in [1,2], \epsilon \in [1,30] $
\ENSURE $\widehat{I}_{GH}(t+h)$ with $t \in [T^*,T]$
\FOR{$n:=1$ to $T^* $}
\IF{$I_{CS}(n)<\epsilon$}
\STATE{$\kappa(n)=\emptyset$}
\ELSE
\STATE $\kappa(n) \leftarrow I_{GH}(n)/I_{CS}(n)$
\ENDIF
\ENDFOR
\STATE{$\bar{\kappa} \leftarrow $mean($\kappa(n))$}
\STATE{$\rho\leftarrow $ACF$(\kappa(n),\kappa(n-h))$}
\STATE{$nn \leftarrow 0$}
\REPEAT 
\STATE $nn\leftarrow nn+1$
\UNTIL {$I_{CS}(t-nn) \geq \epsilon $ } 
\STATE {Pred$\kappa \leftarrow min(\rho \times \kappa(t-nn)+(1-\rho)\times \bar{\kappa},\beta)$}
\STATE {Pred $\leftarrow$ Pred$\kappa \times I_{CS}(t+h)$}
\RETURN {Pred}
\end{algorithmic}
\end{algorithm}
 
  \subsection{Exponential Smoothing}
 \label{algo:ES}
 In practice, it is not necessary to calculate the smoothing on all the in-sample data, limiting to a range covering the daily periodicity ($max = 24$ h) or 2 times this ($max = 48$ h) is sufficient to obtain good results (Algorithm \ref{algo4}).

\begin{algorithm}
\caption{ES}
\begin{algorithmic} 
\label{algo4}
\REQUIRE $ I_{CS}, I_{GH},h>0,\beta \in [1,2], \epsilon \in [1,30], max \in [10-48] $
\ENSURE $\widehat{I}_{GH}(t+h)$ with $t \in [T^*,T]$
\FOR{$n:=1$ to $T^* $}
\IF{$I_{CS}(n)<\epsilon$}
\STATE$\kappa(n)=1$
\ELSE
\STATE $\kappa(n) \leftarrow I_{GH}(n)/I_{CS}(n)$
\ENDIF
\ENDFOR
\STATE{$\bar{\kappa} \leftarrow $mean($\kappa(n))$}
\STATE{$\rho\leftarrow $ACF$(\kappa(n),\kappa(n-h))$}
\FOR{$nn:=0$ to $max-1 $}
\IF{$I_{CS}(t-nn)<\epsilon$}
\STATE{$\kappa(t-nn)=1$}
\ELSE
\STATE $\kappa(t-nn) \leftarrow I_{GH}(t-nn)/I_{CS}(t-nn)$
\ENDIF
\ENDFOR
\STATE {Pred$\kappa \leftarrow min(\rho \times \sum_{i=0}^{max-1} (1-\rho)^i \times \kappa(t-i) +\bar{\kappa} \times (1-\rho)^{max},\beta)$}
\STATE {Pred $\leftarrow$ Pred$\kappa \times I_{CS}(t+h)$}
\RETURN {Pred}
\end{algorithmic}
\end{algorithm}
 
  \subsection{Proposed Methodology \textnormal{(ARTU)}}
 \label{algo:METH}
In this version of the code (Algorithm \ref{algo5}), we propose to associate the night hours with a $\kappa$ equal to 1 but another way which is slightly more complex but which gives very good results consists in neglecting the night hours by removing them completely as done in the Algorithm \ref{algo3} ($\mathbf{if}$ $I_{CS}(n)<\epsilon$ $\mathbf{then}$ $\kappa(n)=\emptyset$). The method requires knowledge of $\alpha$ and $K$, which is achieved by interpolating the $\mathcal{M}(R)$ matrices (see \url{https://github.com/cyrilvoyant/ARTU.git}). Knowing the correlation coefficients ($\rho(h)$ and $\rho(2h) $) and the measurement reliability ($R = 0,0.01,0.05,0.1)$ the interpolation allows an estimate of $\alpha$ and $K$ for these three characteristic values.

\begin{algorithm}
\caption{ARTU}
\begin{algorithmic} 
\label{algo5}
\REQUIRE $ I_{CS}, I_{GH},h>0,\beta \in [1,2], \epsilon \in [1,30], R \in [0,0.01,0.05,0.1], \mathcal{M}(R) $
\ENSURE $\widehat{I}_{GH}(t+h)$ with $t \in [T^*,T]$
\FOR{$n:=1$ to $T^* $}
\IF{$I_{CS}(n)<\epsilon$}
\STATE$\kappa(n)=1$
\ELSE
\STATE $\kappa(n) \leftarrow I_{GH}(n)/I_{CS}(n)$
\ENDIF
\ENDFOR
\STATE{$\bar{\kappa} \leftarrow $mean($\kappa(n))$}
\STATE{$\rho1\leftarrow $ACF$(\kappa(n),\kappa(n-h))$}
\STATE{$\rho2\leftarrow $ACF$(\kappa(n),\kappa(n-2h))$}
\STATE{$(\alpha,K) \leftarrow interpolate(\mathcal{M}(R),\rho1,\rho2,R)$}
\STATE{$S\leftarrow \alpha+K$}
\STATE{$P\leftarrow \alpha \times K$}
\FOR{$nn:=0$ to $h $}
\IF{$I_{CS}(t-nn)<\epsilon$}
\STATE{$\kappa(t-nn)=1$}
\ELSE
\STATE $\kappa(t-nn) \leftarrow I_{GH}(t-nn)/I_{CS}(t-nn)$
\ENDIF
\ENDFOR
\STATE {Pred$\kappa \leftarrow min(S \times \kappa(t)-P \times \kappa(t-h) + (1+P-S) \times \bar{\kappa},\beta)$}
\STATE {Pred $\leftarrow$ Pred$\kappa \times I_{CS}(t+h)$}
\RETURN {Pred}
\end{algorithmic}
\end{algorithm}



\end{document}